\documentclass[11pt, a4paper]{article} 

\usepackage[margin=2.5cm]{geometry}

\usepackage{graphicx}  
\usepackage{float}     
\usepackage{caption}   

\usepackage{amsmath, amssymb, amsfonts}
\usepackage{bm} 

\usepackage[colorlinks=true, linkcolor=blue, citecolor=blue, urlcolor=blue]{hyperref}

\usepackage{authblk}

\title{\textbf{Riemannian vs. Euclidean Representation of Gait Kinematics: A Comparative Analysis}}

\author{Tomáš Bůžek}
\affil{Masaryk University, Faculty of Sports Studies, Brno, Czech Republic}
\date{\today} 

\begin{document}

\maketitle

\begin{abstract}
Accurate quantification of complex human movements, such as gait, is essential for clinical diagnosis and rehabilitation but is often limited by traditional linear models rooted in Euclidean geometry. These frameworks frequently fail to capture the intrinsic non-linear dynamics and posture-dependent dependencies of biological systems. To address this, we present a computational framework that maps kinematic data onto a Riemannian manifold of Symmetric Positive Definite (SPD) matrices. Using the Log-Euclidean metric, we transformed raw skeletal pose sequences into geometric feature vectors to quantify gait variability and smoothness across three velocity profiles: slow, medium, and fast.

Our comparative analysis reveals a critical divergence between geometric approaches. While Euclidean metrics exhibit a strictly linear increase in variability with speed (Slow $<$ Medium $<$ Fast), implying instability, the proposed Riemannian metrics reveal a non-linear ``inverted-U'' pattern with varying speeds. Specifically, we observed a stabilization of variance at high speeds (sprinting), suggesting that the motor system optimizes efficiency by adhering to geodesic trajectories of minimum effort. These findings demonstrate that manifold-based representations offer superior sensitivity to biomechanical efficiency compared to standard linear methods, providing a robust foundation for future diagnostic algorithms and explainable machine learning models in clinical biomechanics.
\end{abstract}

\section{Introduction}

Accurate quantification of human movement, particularly in complex activities like gait, is crucial for clinical diagnosis and the development of personalized rehabilitation strategies \cite{xiang_explainable_2025}. Current analytical approaches, including those processing kinematic data derived from video analysis, typically rely on models rooted in Euclidean geometry \cite{slijepcevic_explaining_2021}. However, the dynamics of complex biological systems are fundamentally characterized by non-linear dependencies and intrinsic geometric structures that linear Euclidean frameworks fail to capture adequately \cite{klein_riemannian_2022}. This reliance on flat-space assumptions often results in ML models exhibiting suboptimal prediction accuracy and reduced robustness, contributing to the so-called 'black-box' problem, which remains a critical barrier for the adoption of AI in clinical gait analysis \cite{xiang_recent_2022, zhang_neural_2025, xiang_explainable_2025}.

To address these limitations, we propose a computational approach that translates raw biomechanical data into a Riemannian (non-Euclidean) space. The configuration space of a multi-linked mechanical system, such as the human body, is naturally described as a Riemannian manifold, which intrinsically accounts for non-linear inertial properties \cite{choi_optimal_2024}. Modeling human motion within this structure allows efficient movements to be represented as geodesics—paths of minimum muscular effort or shortest geometric distance on the manifold—rather than straight lines in Euclidean space \cite{biess_computational_2007}.

This paper presents a comparative evaluation of a computational framework designed to quantify gait patterns using Riemannian geometric principles. By analyzing characteristic gait variables within this non-Euclidean space, we aim to derive diagnostic metrics that are more geometrically consistent than those derived from conventional methods. This explicit geometric structuring enhances the utility of the data for downstream applications, promising superior predictive performance and enabling better interpretability required for advanced Machine Learning systems \cite{xiang_explainable_2025, zhang_neural_2025}.

The remainder of this paper is organized as follows: Section 2 describes the methodology of mapping factorial biomechanics data to the manifold. Section 3 presents the quantitative comparison of gait metrics across different velocity profiles, and Section 4 discusses the implications for future diagnostic tools.

\section{Methodology}

The proposed computational framework processes biomechanical time-series data to extract both geometric manifold features and conventional kinematic metrics. The pipeline consists of three main stages: data preprocessing, Riemannian manifold mapping, and feature vectorization.

\subsection{Data Acquisition and Preprocessing}
The input data consists of raw skeleton pose sequences stored in JSON format, derived from video analysis. Each frame captures the 2D/3D coordinates of detected anatomical keypoints (e.g., hip, knee, ankle). The continuous time-series data is segmented into individual gait cycles (steps), defined by heel-strike events. For each segment, we construct a data matrix $X \in \mathbb{R}^{T \times N}$, where $T$ represents the number of frames in the cycle and $N$ denotes the number of tracked features (joint angles and coordinates).

\subsection{Riemannian Manifold Mapping}
To capture the intrinsic correlation structure of the movement, we utilize the covariance matrix as a robust descriptor. For a centered data segment $X$, the sample covariance matrix $C$ is computed. Since covariance matrices are symmetric and positive-definite (SPD), they do not lie in a Euclidean vector space but form a Riemannian manifold $\mathcal{M} = Sym_d^+$.

Standard Euclidean operations on SPD matrices are geometrically invalid. Therefore, we employ the Log-Euclidean metric framework \cite{arsigny_log-euclidean_2006} to map these points from the curved manifold to the tangent space at the identity matrix. This is achieved via the matrix logarithm:
\begin{equation}
    L = \log_m(C) = U \log(\Sigma) U^T
\end{equation}
where $C = U \Sigma U^T$ is the eigendecomposition of the covariance matrix. The resulting matrix $L$ resides in a Euclidean tangent space, allowing for the application of standard linear statistical methods.

\subsection{Manifold Visualization}
To enable intuitive visual inspection of the high-dimensional geometric structure, the framework incorporates dimensionality reduction techniques. Specifically, we utilize Uniform Manifold Approximation and Projection (UMAP) to project the vectorized Riemannian features ($v$) into a low-dimensional 2D space while preserving the global topological structure of the data manifold \cite{mcinnes_umap_2018}. This allows for the qualitative assessment of gait separability across different velocity profiles.

\subsection{Feature Extraction and Vectorization}
Since $L$ is a symmetric matrix, it contains redundant information. To create a unique "fingerprint" of the walk cycle suitable for machine learning algorithms, we apply a vectorization operator. We extract the unique elements from the upper triangular part of the log-transformed matrix $L$:
\begin{equation}
    v = \text{vech}(L) = [L_{1,1}, L_{1,2}, \dots, L_{d,d}]^T
\end{equation}
These vector embeddings ($v$) serve as the geometric representation of the gait cycle, preserving the intrinsic Riemannian structure of the original motion data.

\subsection{Kinematic Metrics}
In parallel with the geometric analysis, the system calculates explicit physical gait parameters to serve as a baseline for comparison. These include:
\begin{itemize}
    \item \textbf{Speed:} Calculated from the displacement of the root joint (Center of Mass) over time.
    \item \textbf{Step Length:} Derived from the Euclidean distance between ankle keypoints at heel strike.
    \item \textbf{Joint Angles:} Extracted directly from the coordinate vectors of adjacent skeletal segments.
\end{itemize}
\subsection{Baseline Euclidean Metrics}
To assess the specific contribution of the Riemannian framework, we computed equivalent metrics in the standard Euclidean space directly from the raw coordinate data matrices ($X$):

\begin{itemize}
    \item \textbf{Euclidean Smoothness ($E_{smooth}$):} Calculated as the cumulative sum of the Frobenius norms of frame-to-frame velocity vectors, representing the total path length of the motion in flat space.
    \item \textbf{Euclidean Velocity ($E_{v}$):} Defined as the mean magnitude of the velocity vectors of the Center of Mass across the gait cycle.
    \item \textbf{Euclidean Variance ($E_{var}$):} Computed as the trace of the covariance matrix of the raw data ($Tr(Cov(X))$), representing the total variance (sum of variances of all joints) without considering geometric dependencies.
\end{itemize}
\subsection{Statistical Aggregation}
Since each video recording contains multiple gait cycles (steps), variability exists between individual steps. To ensure the robustness of the resulting metrics against measurement outliers and detection noise, we calculate the feature vector for each detected step individually. The final diagnostic value for a given subject/speed is reported as the median of these values across all validated steps.
\subsection{Data Acquisition and Scope of Analysis}
The input data consists of raw skeleton pose sequences derived from video analysis. For the purpose of this comparative analysis, the scope was restricted to unilateral kinematic data, specifically tracking the right-sided lower extremity joints (right hip, right knee, right ankle). Consequently, the derived geometric features and resulting metrics represent the intrinsic dynamics of the right hemibody. While clinical applications would typically require bilateral assessment to capture gait asymmetry, this focused unilateral approach is sufficient to demonstrate the sensitivity of the proposed Riemannian metrics to changes in gait dynamics.

\section{Results and Discussion}

\subsection{Quantitative Comparison}
The proposed Riemannian framework was evaluated on three distinct gait datasets representing different velocities: Slow (jog), Medium (run), and Fast (sprint). To validate the unique contribution of the manifold-based analysis, we compared the extracted Riemannian features against standard Euclidean kinematic baselines. The results are summarized in Table \ref{tab:comparison}.

\begin{table}[h]
\centering
\caption{Comparison of Geometric (Riemannian) vs. Linear (Euclidean) Gait Metrics across three velocity profiles. Note the non-linear behavior of Riemannian metrics at high speeds.}
\label{tab:comparison}
\begin{tabular}{lccc}
\hline
\textbf{Metric} & \textbf{Slow} & \textbf{Medium} & \textbf{Fast} \\ \hline
\multicolumn{4}{l}{\textit{Euclidean Baseline (Linear Space)}} \\
Euclid Smooth ($E_{smooth}$) & 3.240 & 4.254 & 4.603 \\
Euclid Velocity ($E_{v}$) & 0.032 & 0.043 & 0.046 \\
Euclid Variance ($E_{var}$) & 0.054 & 0.097 & 0.121 \\ \hline
\multicolumn{4}{l}{\textit{Riemannian Manifold (Curved Space)}} \\ Riemann Smooth ($R_{smooth}$) & 461.12 & 481.26 & 476.46 \\
Riemann Velocity ($v_{\mu}$) & 4.611 & 4.813 & 4.765 \\
Riemann Variance ($Var_R$) & 4.827 & 8.594 & 7.658 \\ \hline
\end{tabular}
\end{table}

\begin{figure}[h]
    \centering
    \includegraphics[width=0.85\textwidth]{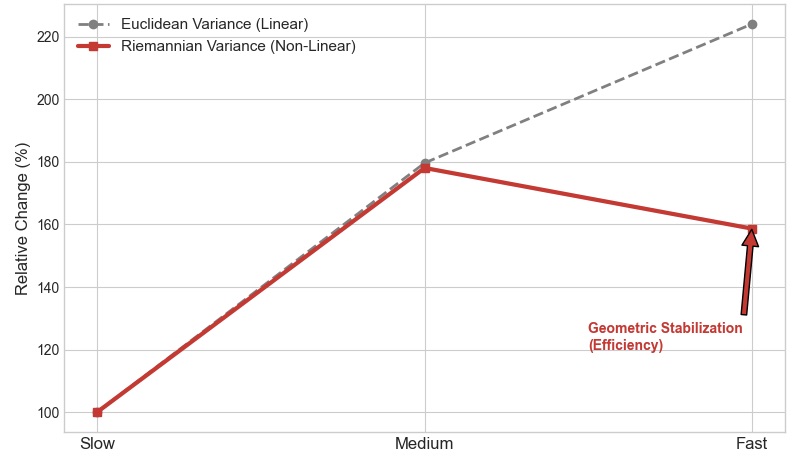}
    \caption{\textbf{Normalized Trend Comparison.} While Euclidean variance (gray, dashed) suggests a linear increase in instability with speed, the proposed Riemannian metric (red, solid) captures the non-linear stabilization at high speeds, reflecting the biomechanical efficiency of the sprint.}
    \label{fig:comparison}
\end{figure}

\subsection{Discussion of Non-Linear Dynamics}
The comparative analysis reveals a fundamental divergence between the two geometric approaches. As hypothesized, the Euclidean metrics exhibit a strictly linear progression (Slow $<$ Medium $<$ Fast). Both path length ($E_{smooth}$) and total variance ($E_{var}$) increase proportionally with speed, suggesting that in Euclidean space, sprinting is interpreted merely as a high-magnitude version of walking. This likely stems from the limitation of linear models to capture posture-dependent non-linearities, leading to an overestimation of variability in complex dynamic tasks \cite{biess_computational_2007, klein_riemannian_2022}.

In contrast, the Riemannian metrics reveal a non-linear relationship, with peak values observed in the Medium velocity profile, followed by a stabilization at Fast velocity. This "inverted-U" pattern supports the hypothesis of \textit{geodesic optimization}. Recent studies suggest that neural dynamics during motor adaptation are constrained by an intrinsic manifold architecture \cite{areshenkoff_neural_2022}. Our observed reduction in Riemannian variability during sprinting aligns with this principle and the theory of Optimal Feedback Control, suggesting that the motor system minimizes "task-irrelevant" noise to strictly adhere to geodesic trajectories when performance demands are highest \cite{biess_computational_2007}.

\section{Conclusion}

This study presented a comparative analysis of human gait kinematics using a Riemannian geometry framework. By transforming raw kinematic data into the space of Symmetric Positive Definite (SPD) matrices, we demonstrated that human motion contains intrinsic geometric structures that are lost in standard Euclidean analysis.

Our experimental results highlight a critical divergence between the two approaches. While Euclidean metrics suggest a linear increase in variability with speed, the Riemannian framework reveals a non-linear stabilization at high velocities. This finding aligns with the theory of optimal feedback control, suggesting that sprinting involves a return to a highly coordinated, geodetically efficient trajectory.

We conclude that the proposed manifold-based representation offers a more physically meaningful baseline for diagnostic algorithms than traditional linear methods. Future work will focus on extending this analysis to bilateral data, which will allow for the precise quantification of gait asymmetry. Such geometric descriptors hold significant potential as robust input features for machine learning models, paving the way for automated pathology detection or post-injury return to load.

\section*{Code Availability}
The source code, including the pipeline for manifold transformation is available at: \url{https://github.com/TommiBu/Riemannian-gait-analysis/tree/main}

\bibliographystyle{unsrt}
\bibliography{references} 

\end{document}